\documentclass[RNAAS]{aastex62}

\begin{document}

\title{Characteristics of the Permanent Superhumps in V533~Herculis}

\correspondingauthor{Peter Garnavich}
\email{pgarnavi@nd.edu}

\author{McKenna Leichty}
\affiliation{Department of Physics and Astronomy, University of Notre Dame, Notre Dame, IN 46556, USA}

\author{Peter Garnavich}
\affiliation{Department of Physics and Astronomy, University of Notre Dame, Notre Dame, IN 46556, USA}

\author{Colin Littlefield}
\affiliation{Department of Physics and Astronomy, University of Notre Dame, Notre Dame, IN 46556, USA}

\author{Rebecca Boyle}
\affiliation{Department of Physics and Astronomy, University of Notre Dame, Notre Dame, IN 46556, USA}

\author{Paul A. Mason}
\affiliation{New Mexico State University, MSC 3DA, Las Cruces, NM, 88003, USA}

\begin{abstract}
We analyze two sectors of TESS photometry of the nova-like cataclysmic variable star V533~Her. We detect a periodicity consistent with the binary orbital period and estimate a revised value of 3.53709(2)~hr. We also detect a strong signal near a period of 3.8~h that we associate with positive superhumps. The superhump frequency varies over the TESS observations with the fractional difference between the superhump and orbital periods, $\epsilon$, ranging between $0.055\le \epsilon \le 0.080$. The superhump amplitude is correlated with its frequency such that the amplitude increases as $\epsilon$ decreases. Positive superhumps result from an instability that generates an eccentric accretion disk and $\epsilon$ is a measure of the disk precession rate in the binary rest frame. The observed correlation implies that as the disk precession rate slows, the disk eccentricity increases.

\end{abstract}


\section{Introduction} 

Accretion disks in cataclysmic variable binaries (CV) can suffer dynamical instabilities that are seen photometrically as periodic oscillations with frequencies close to that of the system's orbital period. These brightness variations are often called superhumps (SH) because they are commonly observed during superoutbursts of SU~UMa type CV \citep{warner85,osaki89}. The dynamical instabilities are excited by the large mass ratios found in CV systems with periods less than 2-hours, but SH are sometimes detected in long-period nova-like CVs and are referred to as ``permanent'' SH \citep{patterson99}. 

Here, we analyze TESS (Transiting Exoplanet Sky Survey) observations of the old nova V533~Her (Nova Herculis 1963). The 3.53-hr orbital period of this nova-like CV was measured spectroscopically by \citet{thorstensen00}. \citet{patterson79} detected a strong, coherent 63.6~s oscillation from V533~Her, suggesting that it was a DQ~Her type intermediate polar. However, by 1982, the high-frequency signal vanished or weakened below detectability \citep{robinson83, boyle22}. V533~Her is classified as a SW~Sextantis type nova-like CV \citep{thorstensen00,rodriguez02}. A SH detection in V533~Her with a period of 3.79~hr is listed in Table~9 of \citet{patterson05}, but no reference to the original data is given.

\vspace{0.5cm}
\section{Analysis}

TESS observed V533~Her over two visits separated by approximately one year. The first visit was during sector 26 (s26) with a 120s cadence starting June 09, 2020. The second was during sector 40 (s40) using the 20s cadence starting June 24, 2021. 

To search for periodic signals in the TESS light curves, we applied the Lomb-Scargle (L-S) power spectrum algorithm \citep{lomb76,scargle82} implemented in {\tt astropy} \citep{astropy}. Two strong signals were immediately obvious, and the L-S spectra of the two sectors are shown in the left panel of Figure~\ref{fig1}. The higher frequency peak is consistent with the candidate orbital periods from \citet{thorstensen00}, while the signal around 3.8~hr apparently corresponds to a SH periodicity \citep{patterson05}.

\vspace{0.5cm}
\subsection{Orbital Period}

Combining the light curves from the two TESS sectors observed  a year apart, allows a precise estimate of the orbital period. However, the L-S spectrum results in two peaks with nearly the same amplitude at 6.78524(4)~c/d and 6.78784(4)~c/d. \citet{thorstensen00} had also narrowed the orbital period to two possibilities: 6.8074~c/d and 6.7843~c/d. Of our choice of periods, the lower frequency candidate is very close to the \citet{thorstensen00} low-frequency candidate, and we conclude that the orbital period of V533~Her is 3.53709(2)~hr.

\subsection{Superhump Oscillations}

\begin{figure}
\begin{center}
\includegraphics[scale=0.43,angle=0]{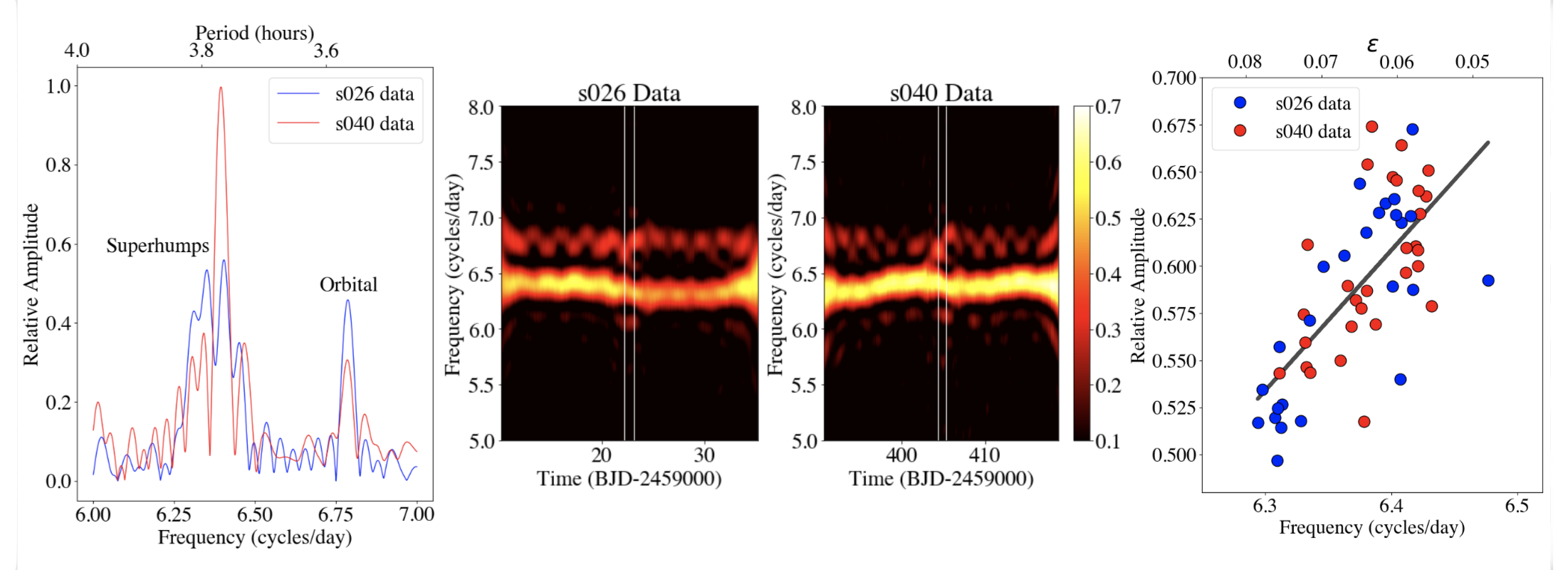}
\caption{{\bf Left:} Amplitude spectra of V533~Her from full TESS light curves for each sector. A strong superhump signal is seen varying around a period of 3.8 hours and a weaker orbital peak is detected with a period of 3.537 hours. {\bf Center:} The dynamical spectra using a moving 5~d window applied to the TESS data. A varying superhump frequency is evident. The apparent oscillations in the orbital period are an artifact resulting from interference with the stronger superhump signal. The vertical white lines on each plot indicate the data download gap in TESS light curves. {\bf Right:} The superhump peak amplitude is plotted against its frequency for both sectors, clearly showing a positive correlation between frequency and amplitude. A linear fit to the points is shown as a solid line with a slope of 0.74 $\pm$ 0.11 in units of amplitude/(cycles~d$^{-1}$). The fractional difference between the superhump and orbital frequencies ($\epsilon$) is displayed at the top of the plot. The value of $\epsilon$ in V533~Her varies by 50\% , suggesting that long-term monitoring is needed to fully characterize the superhump properties in long orbital period systems.   \label{fig1}}
\end{center}
\end{figure}

The strong $\approx$3.8~hr signal in the power spectra appears to correspond to superhump oscillations noted in \citet{patterson05}. These ``permanent'' superhumps are sometimes seen in high mass-transfer rate nova-like CV caused by the excitation of a disk instability \citep{patterson99}. The central panels in Figure~\ref{fig1} display the dynamic L-S spectra for V533~Her using a 5-day window function and clearly shows variability in the disk instability frequency. The fractional difference between the SH and orbital periods is often called $\epsilon$, where $\epsilon=(P_{SH}-P_{orb})/P_{orb}$. Here, $P_{orb} < P_{SH}$, so these are positive SH. Negative SH are not detected in the TESS light curves. The $\epsilon$ value is highly variable in V533~Her, shifting between 0.055 and 0.080 over several weeks. \citet{patterson05} estimated $\epsilon = 0.0719\pm 0.002$, which is at the higher end of our observed range. SH periods are a good indicator of the binary mass ratio in short period systems (e.g. SU~UMa stars), but the compilations by \citet{patterson99} and \citet{patterson05} show a large scatter in $\epsilon$ for the longest-period superhumpers. If V533~Her is typical, the scatter seen in $\epsilon$ for nova-likes with orbital periods $>3$~h may be due to frequency variability on a time-scale of days.

The dynamic L-S spectra suggest that the SH amplitude is smallest when its frequency is low. To test this, we measured the SH peak amplitude and frequency for each step of the moving light curve window and plot the result in the right panel of Figure~\ref{fig1}. The SH amplitude shows a strong correlation with frequency (and $\epsilon$) in the sense that the SH amplitude increases as its frequency approaches the orbital frequency. The average brightness of the star is fairly constant over the individual sectors and there is no detectable correlation between the SH amplitude and the average system flux within the moving window. We therefore directly associate the amplitude of positive SH with the eccentricity of the disk, and we find that the disk becomes more eccentric as $\epsilon$ decreases. 

\vspace{0.5cm}
\section{Discussion}

The dynamical instability seen photometrically as superhumps is believed to be excited by a 3:1 resonance between the outer accretion disk and the orbital period of system \citep{hirose90}. For the disk to extend as far as the 3:1 resonance requires the binary component mass ratio, $M_2/M_1=q$, to be quite small, with $q\lessapprox 0.3$ \citep[e.g.][]{littlefield18}. This naturally explains why SH are commonly detected in CVs with orbital periods below the period gap as these systems typically have secondary star masses of 0.20~M$_\odot$ or less \citep{knigge11}. In contrast, CVs above the period gap will tend to exceed the $q<0.3$ limit as the average white dwarf mass is between 0.7 and 0.8~M$_\odot$ and the secondaries tend to have masses greater than 0.20~M$_\odot$. Given its 3.53~hour orbital period, the secondary in V533~Her is expected to have a mass between 0.25 and 0.3~M$_\odot$. However, if the white dwarf in V533~Her were more massive than average, then it could reach the critical mass ratio. Indeed, based on far-UV spectra, \citet{sion17} estimated the white dwarf in V533~Her to have a mass near 1.0~M$_\odot$, so that the binary mass ratio is expected to be $0.25<q<0.30$. Thus, the presence of positive SH in V533~Her is consistent with the excitation of the 3:1 resonance in the outer disk.

\section{Conclusions}

Based on the TESS observations and the period constraints of \citet{thorstensen00}, we measure an orbital period of 3.53709(2)~hr, the most precise period currently known for this system.

We confirm the presence of positive SH in V533~Her based on two sectors of TESS observations. The SH amplitude is significantly larger than the orbital modulation. The frequency of the superhump oscillations varies between 6.30~c/d and 6.45~c/d over the TESS observations. Using our estimate of the orbital period, we find that $\epsilon=P_{SH}/P_{orb}-1$ meanders between 0.055 and 0.080. This range is surprisingly large and implies that superhumpers with orbital periods greater than 3~hr may have SH periods that are variable over time scales of days. This may explain the wide scatter of $\epsilon$ values measured for long-period systems in \citet{patterson99, patterson05}. Certainly, caution is in order when estimating the binary mass ratio, $q$, from $\epsilon$ measurements in long-period superhumpers.

We identify a strong correlation between the SH amplitude and frequency in V533~Her. If positive SH result from a precessing eccentric accretion disk, then this correlation implies that the slower the precession rate in the binary restframe, the greater the eccentricity of the disk.

\acknowledgments

We acknowledge support from TESS/NASA grant 80NSSC22K0183.


\begin{thebibliography}{}

\bibitem[Astropy Collaboration et al.(2013)]{astropy} Astropy Collaboration, Robitaille, T.~P., Tollerud, E.~J., et al.\ 2013, \aap, 558, A33. doi:10.1051/0004-6361/201322068


\bibitem[Boyle et al.(2022)]{boyle22} Boyle, R.\ 2022, in prep.


\bibitem[Hirose \& Osaki(1990)]{hirose90} Hirose, M. \& Osaki, Y.\ 1990, \pasj, 42, 135

\bibitem[Knigge et al.(2011)]{knigge11} Knigge, C., Baraffe, I., \& Patterson, J.\ 2011, \apjs, 194, 28. doi:10.1088/0067-0049/194/2/28


\bibitem[Littlefield et al.(2018)]{littlefield18} Littlefield, C., Garnavich, P., Kennedy, M., et al.\ 2018, \aj, 155, 232. doi:10.3847/1538-3881/aabcd1


\bibitem[Lomb(1976)]{lomb76} Lomb, N.~R.\ 1976, \apss, 39, 447. doi:10.1007/BF00648343

\bibitem[Osaki(1989)]{osaki89} Osaki, Y.\ 1989, \pasj, 41, 1005


\bibitem[Patterson(1979)]{patterson79} Patterson, J.\ 1979, \apjl, 233, L13. doi:10.1086/183066

\bibitem[Patterson(1999)]{patterson99} Patterson, J.\ 1999, Disk Instabilities in Close Binary Systems, 61

\bibitem[Patterson et al.(2005)]{patterson05} Patterson, J., Kemp, J., Harvey, D.~A., et al.\ 2005, \pasp, 117, 1204. doi:10.1086/447771

\bibitem[Robinson \& Nather(1983)]{robinson83} Robinson, E.~L. \& Nather, R.~E.\ 1983, \apj, 273, 255. doi:10.1086/161364

\bibitem[Rodr{\'\i}guez-Gil \& Mart{\'\i}nez-Pais(2002)]{rodriguez02} Rodr{\'\i}guez-Gil, P. \& Mart{\'\i}nez-Pais, I.~G.\ 2002, Classical Nova Explosions, 637, 558. doi:10.1063/1.1518263


\bibitem[Scargle(1982)]{scargle82} Scargle, J.~D.\ 1982, \apj, 263, 835. doi:10.1086/160554

\bibitem[Sion et al.(2017)]{sion17} Sion, E.~M., Godon, P., \& Jones, L.\ 2017, \aj, 153, 109. doi:10.3847/1538-3881/153/3/109


\bibitem[Thorstensen \& Taylor(2000)]{thorstensen00} Thorstensen, J.~R. \& Taylor, C.~J.\ 2000, \mnras, 312, 629. doi:10.1046/j.1365-8711.2000.03230.x


\bibitem[Warner(1985)]{warner85} Warner, B.\ 1985, Interacting Binaries, 150, 367


\end{thebibliography}
\end{document}